\documentstyle[aps,pre]{revtex}

\begin{document}
\draft \title{Multiphonon and ``hot''-phonon Isovector Electric-Dipole
Excitations} \author{B.V. Carlson$^1$, L.F. Canto$^2$,
S. Cruz-Barrios$^3$, M.S. Hussein$^4$, and A.F.R. de Toledo Piza$^4$}
\address{$^1$Departamento de F\'{\i}sica, Instituto Tecnol\'{o}gico
da\\ Aeron\'{a}utica -- CTA, 12228-900 S\~{a}o Jos\'{e} dos Campos SP
Brazil\\ $^2$Instituto de F\'{\i}sica da Universidade do Rio de
Janeiro, CP 68528,\\ 21945-970 Rio de Janeiro RJ, Brazil\\
$^3$Departamentode F\'{\i}sica Aplicada, Universidad de Sevilla\\41080
Sevilla, Spain \\ 
$^4$Instituto de F\'{\i}sica, Universidade de S\~{a}o Paulo, CP
66318,\\ 05389-970 S\~{a}o Paulo SP, Brazil}
\date{\today}

\maketitle

\begin{abstract}

We argue that a substantial increase in the cross section for Coulomb
excitation in the region of the Double Giant Dipole Resonance should
be expected from Coulomb excitation of excited states involved in the
spreading of the one-phonon resonance, in a manifestation of the
Brink-Axel phenomenon. This generates an additional fluctuating amplitude
and a corresponding new term to be added incoherently to the usual
cross-section. The appropriate extension of an applicable reaction
calculation is considered in order to estimate this effect.
\end{abstract}
\vspace{0.25 cm}


\vspace{0.25 cm}


The Coulomb excitation of two-phonon giant resonances at intermediate
energies has generated considerable interest in the last few
years\cite {emling}. The isovector double giant dipole resonance
(DGDR) has been observed in $^{136}$Xe\cite{schmidt},
$^{197}$Au\cite{aumann}, and $^{208}$Pb
\cite{ritman,beene,boretzky}. The isoscalar double giant quadrupole
resonance has also been observed in the proton emission spectrum from
the collision of $^{40}$Ca with $^{40}$Ca at a laboratory energy of 44
A Mev\cite{chomaz}.When the data on DGDR excitation for $^{136}$Xe and
$^{197}$ Au are compared with coupled-channel Coulomb excitation
calculations\cite {canto}, it is found that, in the harmonic
approximation, the calculated cross sections are a factor of 2 to 3
smaller than the measured ones. A similar discrepancy, albeit somewhat
smaller, is found for $^{208}$Pb.

Several effects that are not taken into account in the coupled-channel
theory have been considered as possible explanations of this
discrepancy. As examples, we mention the effect of
anharmonicities\cite{volpe,bortignon} and the quenching of the $1^{+}$
DGDR state\cite{bertu}. Here we will consider a new potentially
important mechanism, which consists in the (one phonon) Coulomb
excitation of backgrount states responsible for the large spreading
width of the one-phonon GDR, as suggested long ago by Brink and
Axel\cite{BrAx}. Due to the complicated background of intrinsic
states, the amplitude for this process varies rapidly wth energy and
possesses an average close to zero. Its contribution to the cross
section can be sizable, however. In close analogy to this situation is
the the well-known case of nucleon-nucleus elastic scattering. There,
the cross section is the sum of the slowly-varying contribution of
average optical scattering and of the fluctuating contribution
compound nucleus formation and decay. In figure 1 we show a schematic
picture of the couplings involved.

We first sumarize our main result. The cross-section for Coulomb
excitation to the DGDR energy region contains in fact two distinct
components which peak at $\sim 2E_{GDR}$. However, while the usual
component $\sigma_{DGDR}$ has a width which may be estimated as $\sim
2\Gamma_{GDR}$, the fluctuating Brink-Axel component has a width which
is just $\sim \Gamma_{GDR}$. As a result of this, the bump observed in
the two-phonon region has an effective width between these limits. The
enhancement factor for the peak-value of the cross-section will be
given roughly by $(1+\Gamma_1^{\downarrow}/\Gamma_1)$, whereas the
cross-section integrated over the peak is just about
$(1+\frac{1}{2}\Gamma_1^{\downarrow}/\Gamma_1)\sigma_{DGDR}$ due to
the smaller width of the second component. For heavy nuclei
$\Gamma_1^{\downarrow}\sim\Gamma_1$, and we get enhancement factors of
$\sim 2$ and $\sim 3/2$ for the peak and for the integrated
cross-sections respectively. Furthermore, these enhancement factors
should be reduced and tend to unity as the collision time becomes
shorter than the inverse $GDR$ width, at higher bombarding energies. A
simple model illustrating these features will be discussed in what
follows.

Giant resonances have many characteristics that suggest a treatment in
terms of simple collective degrees of freedom. The first and foremost
of these is their classical interpretation in terms of macroscopic
shape oscillations of the nucleus. The properties of multiple
excitations of these resonances would then suggest that they are
simple bosonic degrees of freedom. The Brink-Axel hypothesis, which
assumes that a giant dipole resonance may be constructed on each of
the intrinsic excited states of the nucleus, suggests that the
resonances can be considered as degrees of freedom independent of the
intrinsic states. Of course, the microscopic representation of the
giant resonances, in terms of of the intrinsic particle-hole states,
implies that their treatment as independent degrees of freedom can
only be approximate.  Yet, in many instances, it seems to be a very
good approximation.

As a model of the multiple excitation and decay of a giant resonance,
we will consider the excitation of a collective degree of freedom of a
target nucleus by an inert projectile and the subsequent decay of the
collective states into complex intrinsic ones. We can loosely follow
the development given by Ko years ago\cite{ko}. His work was motivated
by the need to incorporate the collective features of the Copenhagen
approach to deeply-inelastic heavy-ion collisions \cite{broglia} into
the Heidelberg description \cite{aga} of these processes.

As did Ko, we take the Hamiltonian describing the two colliding nuclei
to be
\begin{equation}
H(\vec{r},\alpha ,\xi )=h_0(\vec{r})+h_1(\alpha )+h_2(\xi )+U(\vec{r},\xi
)+V(\vec{r},\alpha )+W(\alpha ,\xi )  \label{ham}
\end{equation}
where $h_0(\vec{r})=\vec{p}^2/2m+V_0(\vec{r})$ is the Hamiltonian for
the relative motion in the ground state, $h_1(\alpha )$ and $h_2(\xi
)$ are the collective and intrinsic Hamiltonians, respectively,
$W(\alpha ,\xi )$ is the coupling between the collective and the
intrinsic degrees of freedom and $U(\vec{r},\xi )$ and
$V(\vec{r},\alpha )$ are the couplings between the relative coordinate
and the intrinsic and collective degrees of freedom, respectively. As
we wish to concentrate our attention on the excitation of the
collective states alone through the relative motion, we simplify by
taking the direct coupling of the relative motion to the intrinsic
states to be null, $U(\vec{r},\xi )=0$. We write the collective and
intrinsic spectra and states as $h_1(\alpha )\left| n\right\rangle
=e_n\left| n\right\rangle$ and $h_2(\xi )\left| \nu \right\rangle
=\varepsilon _\nu \left| \nu\right\rangle$. In the case that the
collective spectrum represents multiple excitations of a giant
resonance, we would expect $e_n\simeq n\,e_1.$

We write the uncoupled scattering states of the relative motion with
an incoming or outgoing wave boundary condition corresponding to
asymptotic wavenumber $\vec{k}$ as
\begin{equation}
h_0\ \left| \varphi _{\vec{k}}^{\pm }\right\rangle =\left( \frac{\ \ \vec{p}%
^{\ 2}}{2m}+V_0\right) \ \left| \varphi _{\vec{k}}^{\pm }\right\rangle =E_{%
\vec{k}}\ \left| \varphi _{\vec{k}}^{\pm }\right\rangle .
\label{rspec}
\end{equation}
The free Green's function can then be expressed in terms of a diagonal
sum over the triple product states as
\begin{equation}
G_0^{+}(E)=\,\sum_{n,\nu }\,\int \,d^3k\,\frac{\left| n,\nu ,\varphi _{\vec{k%
}}^{+}\right\rangle \left\langle n,\nu ,\varphi _{\vec{k}}^{+}\right| }{E-E_{%
\vec{k}}-e_n-\varepsilon _\nu +i\,\eta }. \label{green0}
\end{equation}
We point out that an equivalent expression for $G_0^{+}(E)\,$ is
obtained by substituting the scattering states satisfying an outgoing
wave boundary conditions for the incoming wave states used above.

In this basis, the matrix elements of the couplings take the form
\begin{equation}
\left\langle m,\mu ,\varphi _{\vec{k}}^{+}\right| V(\vec{r},\alpha
)\left| n,\nu ,\varphi _{\vec{k}^{\prime }}^{+}\right\rangle =\delta
_{\mu \nu
}\,\left\langle m,\varphi _{\vec{k}}^{+}\right| V\,\left| n,\varphi _{\vec{k}%
^{\prime }}^{+}\right\rangle \label{v2del}
\end{equation}
$^{}$and 
\begin{equation}
\left\langle m,\mu ,\varphi _{\vec{k}}^{+}\right| W(\alpha ,\xi )\left|
n,\nu ,\varphi _{\vec{k}^{\prime }}^{+}\right\rangle =\delta ^{(3)}(\vec{k}-%
\vec{k}^{\prime })\,\,\left\langle m,\mu \right| W\,\left| n,\nu
\right\rangle .  \label{v3del}
\end{equation}
The first of these says that the coupling of the relative motion to
the collective degree of freedom due to $V(\vec{r},\alpha )$ does not
affect the intrinsic state. The second says that the coupling of the
collective degree of freedom to the intrinsic ones due to $W(\alpha
,\xi )$ does not affect the relative motion of the projectile and
target. The transitions induced by $W(\alpha ,\xi )$ are internal to
the target.

We assume that the complex intrinsic states are statistical and use a
schematic random-matrix model to describe their matrix elements.We
take for the first and second moments of the matrix elements
$\overline{\left\langle m,\mu \right| W\,\left| n,\nu
\right\rangle}=0$
and 
\begin{equation}
\,\overline{\left\langle m,\mu \right| W\,\left| n,\nu \right\rangle
\,\left\langle n^{\prime },\nu ^{\prime }\right| W\,\left| m^{\prime },\mu
^{\prime }\right\rangle }=\delta _{mm^{\prime }}\delta _{nn^{\prime }}\delta
_{\mu \mu ^{\prime }}\delta _{\nu \nu ^{\prime }}\,\overline{\left|
\left\langle n,\nu \right| W\,\left| m,\mu \right\rangle \right| ^2}
\label{mat2}
\end{equation}
The statistical hypotheses only require that the average squared
matrix elements vanish for $\mu \neq \mu ^{\prime }$ or $\nu \neq \nu
^{\prime }$.  For simplicity, we take them to vanish for $m\neq
m^{\prime }$ and $n\neq
n^{\prime }$ as well.

We can now use the statistical hypotheses on the matrix elements of
$W$ to calculate the average (optical) Green's function
$\overline{G}^{+}(E)$. This has been done by Ko \cite{ko} and his
result can be written as (see also \cite{fus})
\begin{equation}
\overline{G}_{\vec{k}n\nu }^{+}(E)=\frac 1{E-E_{\vec{k}}-e_n-\varepsilon
_\nu +i\,\Gamma _{n\nu }\,/2\,}  \label{gbarelt1}
\end{equation}
where $\Gamma_{n\nu}$ is the total width of the resonance, comprising
an escape width $\Gamma_{n\nu}^{\uparrow}$ plus a spreading width
$\Gamma_{n\nu}^{\downarrow}$, viz. $\Gamma _{n\nu }=
\Gamma_{n\nu}^{\downarrow }+\Gamma _{n\nu }^{\uparrow}$.  The complete
expression for the average Green's function is then
\begin{equation}
\overline{G}^{+}(E)=\,\sum_{n,\nu }\,\int \,d^3k\,\frac{\left| n,\nu
,\varphi _{\vec{k}}^{+}\right\rangle \left\langle n,\nu ,\varphi _{\vec{k}%
}^{+}\right| }{E-E_{\vec{k}}-e_n-\varepsilon _\nu +i\,\Gamma _{n\nu }\,/2}
\label{avgrn}
\end{equation}
It is worthwhile pausing a moment to interpret this expression. With
the exception of the ground-state component of the Green's function,
all others have a finite width in their denominator, However, the
scattering states that enter are still those corresponding to the
self-adjoint Hamiltonian $h_0$. Yet, in contrast to the free Green's
function, for which the component corresponding to the
collective-intrinsic state $\left| n,\nu\right\rangle $ contains only
the scattering states $\left| \varphi _{\vec{k}}^{+}\right\rangle $
with energy $E_{\vec{k}}=E-e_n-\varepsilon _\nu $, the component of
the average Green's function contains an envelope of scattering states
about this energy, with the extent of the envelope determined by the
width $\Gamma _{n\nu }\,.$ The relative phases of the contributions to
this envelope are such that the outgoing waves of $\overline{G}^{+}$
are decaying waves, as can easily be verified by evaluating the
integral as a contour integral in the complex plane.

We note that at low excitation energies, the widths of the $\left|
1,\nu \right\rangle $ states are dominated by their spreading widths,
as the contributions to the escape widths, from both the collective
state and the intrinsic states, are small. The states thus appear to
be consistent with the Brink-Axel hypothesis, by which the same
collective state (with the same width) is constructed on each of the
intrinsic states. However, as the excitation energy of the intrinsic
states increases, a corresponding increase in both the spreading width
and the escape width of the state becomes observable, consistent with
the increase in widths experimentally observed in hot giant
resonances. Using a phenomenological expression for the compound
escape width, $\Gamma _\nu \approx 14\exp (-4.69\sqrt{A/E^{*}})$ MeV
and taking Sn as an example, we expect the total width to saturate at
high excitation at $\Gamma _{1\nu }\approx $ 4.5 MeV + 14 MeV, which
compares fairly well with the observed value of 15
MeV\cite{gaardhoje}.

The amplitude of the first excited collective state is obtained from a
single action of the coupling $V$. Taking the final relative momentum
of the projectile and target to be $\vec{k}^{\prime }$, the amplitude
$A_1(\vec{k},\vec{k}^{\prime })$ for excitation of the first
collective state is
\begin{equation}
A_1(\vec{k},\vec{k}^{\prime })=\frac 1{E-E_{\vec{k}^{\prime }}-e_1+i\,\Gamma
_{10}\,/2\,}\left\langle 1,\varphi _{\vec{k}^{\prime }}^{-}\right| V\,\left|
\psi _{\vec{k}00}^{+}\right\rangle,  \label{amp1}
\end{equation}
where $\left|\psi_{\vec{k}00}^{+}\right\rangle$ is the relative motion
wavefunction in the entrance channel calculated by taking into account
the coupling to the one-phonon excited state to all orders \cite{canto}.
What is observed are the decay products of the excited state, which can
decay either directly or after passing through the intrinsic states. The
direct contribution is 
\begin{equation}
d\sigma _1^{dir}(\vec{k},\vec{k}^{\prime })=\Gamma _{10}^{\uparrow }\,\left|
A_1(\vec{k},\vec{k}^{\prime })\,\right| ^2,  \label{dsig1d}
\end{equation}
while the decay through the intrinsic states yields 
\begin{equation}
d\sigma _1^{int}(\vec{k},\vec{k}^{\prime })=\Gamma _{01}\frac{\Gamma
_{10}^{\downarrow }}{\Gamma _{01}}\,\left| A_1(\vec{k},\vec{k}^{\prime
})\,\right| ^2=\Gamma _{10}^{\downarrow }\,\left| A_1(\vec{k},\vec{k}%
^{\prime })\,\right| ^2,  \label{dsig1c}
\end{equation}
where $\Gamma _{01}$ is the width of the intrinsic states at an energy $%
\varepsilon _\nu \approx e_1$.We have assumed, based on the complexity of
the intrinsic states, that these possess no spreading width so that the
width $\Gamma _{01}$ is all escape width. Adding the two contributions, we
have for the cross section 
\begin{equation}
d\sigma _1(\vec{k},\vec{k}^{\prime })=\frac{\Gamma _{10}}{(E-E_{\vec{k}%
^{\prime }}-e_1)^2+(\Gamma _{10}\,/2)^2\,}\left| \left\langle 1,\varphi _{%
\vec{k}^{\prime }}^{-}\right| V\,\left| \psi _{\vec{k}00}^{+}\right\rangle
\right| ^2.  \label{dsig1}
\end{equation}

The second collective state can be populated predominantly through a
two-step process. Assuming a final relative momentum of
$\vec{k}^{\prime \prime }$, we have
\begin{eqnarray}
A_2(\vec{k},\vec{k}^{\prime \prime }) &=&\frac 1{E-E_{\vec{k}^{\prime
\prime }}-e_2+i\,\Gamma _{20}\,/2\,}\times \\ & & \int
d^{3\,}k^{\prime }\left\langle 2,\varphi _{\vec{k}^{\prime \prime
}}^{-}\right| V\,\left| 1,\varphi _{\vec{k}^{\prime
}}^{+}\right\rangle \frac 1{E-E_{\vec{k}^{\prime }}-e_1+i\,\Gamma
_{10}\,/2\,}\left\langle
1,\varphi _{\vec{k}^{\prime }}^{+}\right| V\,\left| \psi _{\vec{k}%
00}^{+}\right\rangle  \nonumber
\end{eqnarray}
This amplitude describes the process in which a second collective excitation
occurs before the first collective state has decayed to the intrinsic states.
The corresponding cross section is 
\begin{eqnarray}
d\sigma _2(\vec{k},\vec{k}^{\prime \prime }) &=&\frac{\Gamma _{20}}{(E-E_{%
\vec{k}^{\prime \prime }}-e_2)^2+(\Gamma _{20}\,/2)^2\,}\times
\label{dsig2} \\ &&\left| \int d^{3\,}k^{\prime }\left\langle
2,\varphi _{\vec{k}^{\prime \prime }}^{-}\right| V\,\left| 1,\varphi
_{\vec{k}^{\prime
}}^{+}\right\rangle \frac 1{E-E_{\vec{k}^{\prime }}-e_1+i\,\Gamma _{10}\,/2\,%
}\left\langle 1,\varphi _{\vec{k}^{\prime }}^{+}\right| V\,\left| \psi _{%
\vec{k}00}^{+}\right\rangle \right| ^2  \nonumber
\end{eqnarray}
For a harmonic mode, we expect $e_2\approx 2\,e_1.$We also expect $\Gamma
_{20}\approx 2\,\Gamma _{10}$, since we expect that $\left\langle 2\left|
W\,\right| 1\right\rangle \approx \sqrt{2}\,\left\langle 1\left| W\,\right|
0\right\rangle $.

There is another two-step process -- a fluctuating one -- that can
look as if it were an excitation of the second collective state,
although in fact it is not. This is the process in which a second collective
excitation occurs on top of the hot background of incoherent, intrinsic
excitations remaining after the first collective state has decayed to the
intrinsic states. The amplitude for this process, for an arbitrary
intrinsic state $\left| \nu \right\rangle$, is
\begin{eqnarray}
A_{2,\nu}^{\rm fl}(\vec{k},\vec{k}^{\prime \prime }) &=&\frac
1{E-E_{\vec{k}^{\prime \prime }}-e_1-\varepsilon _\nu +i\Gamma
_{1\nu }/2}\int d^{3}k^{\prime }\left\langle 1\varphi
_{\vec{k}^{\prime \prime }}^{-}\right| V\left| 0\varphi
_{\vec{k}^{\prime }}^{+}\right\rangle
\label{amp2p}\times \\
&&\frac 1{E-E_{\vec{k}^{\prime }}-\varepsilon _\nu +i\Gamma _{0\nu
}/2}\left\langle 0\nu \left| W\right| 10\right\rangle \frac 1{E-E_{%
\vec{k}^{\prime }}-e_1+i\Gamma _{10}/2}\left\langle 1\varphi _{\vec{k}%
^{\prime }}^{+}\right| V\left| \psi _{\vec{k}00}^{+}\right\rangle . 
\nonumber
\end{eqnarray}
Its contribution to the cross section is incoherent with the others. When
summed over the intrinsic states, it is 
\begin{eqnarray}
d\sigma _2^{\rm fl}(\vec{k},\vec{k}^{\prime \prime })=\sum_\nu
\left|A^{\rm fl}_{2,\nu}(\vec{k},\vec{k}^{\prime\prime})\right|^2.
\end{eqnarray}

To get even a crude estimate of the above expression, we have to do a bit of
hand waving. To begin, we replace the intrinsic excitation energy $%
\varepsilon _\nu $ and the decay width $\Gamma _{1\nu }$ by their average
values, $e_1$and $\Gamma _{1c}$ in the final-state factor, so that 
\begin{equation}
\frac{\Gamma _{1\nu }}{(E-E_{\vec{k}^{\prime \prime }}-e_1-\varepsilon _\nu
)^2+(\Gamma _{1\nu }\,/2)^2\,}\longrightarrow \frac{\Gamma _{1c}}{(E-E_{\vec{%
k}^{\prime \prime }}-2\,e_1)^2+(\Gamma _{1c}\,/2)^2\,},  \label{pox1}
\end{equation}
and the factor can be removed from the sum. Next, we approximate the
remaining sum over intrinsic states as 
\begin{equation}
\sum_\nu \,\frac{\left| \left\langle 0,\nu \left| W\,\right|
1,0\right\rangle \right| ^2}{(E-E_{\vec{k}^{\prime }}-\varepsilon _\nu
+i\,\Gamma _{0\nu }\,/2)\,(E-E_{\vec{q}^{\prime }}-\varepsilon _\nu
-i\,\Gamma _{0\nu }\,/2)}\approx \frac{\,\Gamma _{10}^{\downarrow }}{\Gamma
_{0c}}\frac{\,1}{1+i\,(E_{\vec{k}^{\prime }}-E_{\vec{q}^{\prime }})/\Gamma
_{0c}},  \label{pox2}
\end{equation}
We have also replaced the intrinsic state decay widths $\Gamma _{0\nu
}$ by their average value $ \Gamma _{0c}$ and have used
$\vec{k}^{\prime }$ and $\vec{q}^{\prime }$ to denote the dummy
variables of the two conjugate integrals. Finally, we argue that the
restriction imposed on the momentum integrals by the right hand side
of Eq.\ref{pox2} reduces them from their unrestricted value by a
factor of $\Gamma _{0c}/\Gamma _{10}$. That is,
\begin{equation}
\int d^3k^{\prime }\int d^3q^{\prime }\frac{\,F(\vec{k}^{\prime })\,F^{*}(%
\vec{q}^{\prime })}{1+i\,(E_{\vec{k}^{\prime }}-E_{\vec{q}^{\prime
}})/\Gamma _{0c}}\approx \frac{\Gamma _{0c}}{\Gamma _{10}}\left| \int
d^3k^{\prime }\,F(\vec{k}^{\prime })\right| ^2,  \label{pox3}
\end{equation}
where $\,F(\vec{k}^{\prime })$ is the rest of the integrand. With these
three approximations, we can rewrite the fluctuation contribution to the
cross section as 
\begin{eqnarray}
d\sigma _2^{\rm fl}(\vec{k},\vec{k}^{\prime \prime }) &\approx &\frac{\Gamma
_{1c}}{(E-E_{\vec{k}^{\prime \prime }}-2\,e_1)^2+(\Gamma _{1c}\,/2)^2\,}%
\frac{\Gamma _{10}^{\downarrow }}{\Gamma _{10}}  \label{dsig2pf}\times \\
&&\left| \int d^{3\,}k^{\prime }\left\langle 1,\varphi _{\vec{k}%
^{\prime \prime }}^{-}\right| V\,\left| 0,\varphi _{\vec{k}^{\prime
}}^{+}\right\rangle \frac 1{E-E_{\vec{k}^{\prime }}-e_1+i\,\Gamma _{10}\,/2\,%
}\left\langle 1,\varphi _{\vec{k}^{\prime }}^{+}\right| V\,\left| \psi _{%
\vec{k}00}^{+}\right\rangle \right| ^2.  \nonumber
\end{eqnarray}

We can now compare the fluctuation contribution to the apparent
excitation of the second collective state with the actual one of
Eq.\ref{dsig2}. We first observe that both will have approximately the
same average excitation energy of $2\,e_1$. The width of the
fluctuation cross section, though, will be about half that of the real
one, $\Gamma _{1c}\approx \Gamma _{10}\approx \Gamma _{20}/2$, since
the contribution of the intrinsic states to the width is expected to
be extremely small at these energies. Due to the difference in the
matrix elements of $V$, the magnitude squared of the momentum integral
of the fluctuation cross section will also be about half that of the
actual one. The two cross sections will then be comparable at their
peak values. The observed cross section, in these conditions, would be
appreciably larger than the expected value and would have a width intermediate
between the width $\Gamma _{10}$ of the single giant resonance and the
width $\Gamma _{20}\approx 2\,\Gamma _{10}$ expected for the double
giant resonance.

Since the fluctuation contribution to the apparent excitation of the
second collective state depends on the decay of the collective state
into the intrinsic states, we would expect it to be important in a
limited range of incident energy. At sufficiently high incident
energies, we expect that the target would no longer have time to decay
and be excited a second time by the projectile. This tendency can be
seen in the experimental data for $ ^{208}$Pb. The DGDR excitation
cross section observed at a laboratory energy of about 100A MeV is a
factor of two larger than the predicted one\cite{beene}, while the
cross section at 640A MeV is only about 30\% greater than that
calculated\cite{boretzky}. The observed width of the DGDR also tends
to increase with the incident energy, consistent with the diminishing
contribution of the fluctuation cross section.

An extension of the semiclassical model of ref. \cite{canto} to
include the Brink-Axel effect will be reported elsewhere.

We acknowledge partial financial support from MCT/FINEP/CNPq (PRONEX)
contract N. 41.96.0886.00, Funda\c{c}\~{a}o de Amparo \`{a} Pesquisa
do Estado do Rio de Janeiro (FAPERJ), Funda\c{c}\~{a}o de Amparo \`{a}
Pesquisa do Estado de S\~{a}o Paulo (FAPESP), contract N. 96/1381-0,
and Funda\c{c}\~{a}o Universit\'{a}ria Jos\'{e} Bonif\'{a}cio. BVC,
LFC and MSH are research fellows of CNPq. SC-B was partly supported by
PRP-USP (Universidade de S\~ao Paulo) and CICYT, project PB95-0533
(Spain).

\pagebreak

{\bf FIGURE CAPTION}

\bigskip

Figure 1: A schematic picture showing the coupling to the one "cold" phonon
state (d), the two "cold" phonons state, the fine structure states of d (b) and
the one "hot" Brink-Axel (B-A) phonon state.

\end{document}